\documentclass[prb,twocolumn,showpacs,noshowkeys,
preprintnumbers,amsmath,amssymb,byrevtex]{revtex4}
\usepackage{graphicx}% Include figure files
\usepackage{dcolumn}% Align table columns on decimal point
\usepackage{bm}% bold math
\usepackage{times}
\begin{document}

\preprint{}

\title{Two-level correlation function of critical random-matrix ensembles}
\author{E. Cuevas}
%%%\email{ecr@um.es}
%%%\homepage{http://bohr.fcu.um.es/miembros/ecr/}
\affiliation{Departamento de F{\'\i}sica, Universidad de Murcia,
E-30071 Murcia, Spain.}

\date{\today}

\begin{abstract}
The two-level correlation function $R_{d,\beta}(s)$ of $d$-dimensional
disordered models ($d=1$, $2$, and $3$) with long-range random-hopping
amplitudes is investigated numerically at criticality. We focus on
models with orthogonal ($\beta=1$) or unitary ($\beta=2$) symmetry in
the strong ($b^d \ll 1$) coupling regime, where the parameter $b^{-d}$
plays the role of the coupling constant of the model. It is found that
$R_{d,\beta}(s)$ is of the form
$R_{d,\beta}(s)=1+\delta(s)-F_{\beta}(s^{\beta}/b^{d\beta})$, where
$F_{1}(x)=\text{erfc}(a_{d,\beta}\,x)$ and $F_{2}(x)=\exp (-a_{d,\beta}\,x^2)$,
with $a_{d,\beta}$ being a numerical coefficient depending on the
dimensionality and the universality class. Finally, the level number
variance and the spectral compressibility are also considerded.
\end{abstract}


%%%\keywords{Suggested keywords}

\maketitle

\section{Introduction}
\label{int}

Random-matrix theories are largely and successfully applied in the
theoretical description of complex nuclei \cite{Wi58,Dy62}, gauge
field theories, \cite{VZ93,BM98,MB99}
mesoscopic systems, \cite{Be97,FS97,GM98,Al00} and random surfaces in
the field of quantum gravity. \cite{Am94} Their advantage is that it
is possible to represent the Hamiltonian of the corresponding system
by a large Hermitian matrix acting on a finite-dimensional Hilbert
space (if one disregards the continuous part of the spectrum).

The random matrices fall into one of three universality classes,
named orthogonal ($\beta=1$), unitary ($\beta=2$), and symplectic
($\beta=4$), depending on the global symmetry properties of the
Hamiltonian they represent. \cite{Me91} The symmetry parameter $\beta$
is the number of independent real components that characterizes a matrix
element of the Hamiltonian. A system belongs to the orthogonal class if
it has both time-reversal and spin-rotation symmetries, to the unitary
class if time-reversal symmetry is broken, and to the symplectic class
if the system has time-reversal symmetry but spin-rotation is broken.
The relevant terms in the Hamiltonian are a coupling to an applied
magnetic field, which breaks time-reversal symmetry, and the spin-orbit
interaction, which breaks spin-rotation symmetry.

One of the most relevant applications of the random matrix ensembles
is to the study of critical phenomena, particularly to the special
case of critical statistics which is found at the Anderson metal-insulator
transition (MIT) in disordered systems. \cite{AZ88,SS93,KL94,AS86,AK94,%
AM95,Ni99,Ev94,VH95,BM95,GJ98,JK99,BG99,AJ01} At the critical point, the states
acquire the property of multifractality, which marks a qualitative difference
from the extended states in a metal and localized states in an insulator.
These critical states correspond to critical-level statistics. Although
several ensembles of nonconventional random matrices have been suggested
to describe this statistics \cite{KM97,MN94,MC93,GV03,BB97,GV00,BC94} we
wish to emphasize the power-law random-banded matrix model (PRBM), \cite{MF96}
for which the multifractality of eigenstates has been rigorously proven.
\cite{MF96,Mi00} This model is characterized by a variance of their off-diagonal
matrix elements, which decay as a power law with increasing distance from the
diagonal. It should be mentioned that all these models are of a one-dimensional
nature.

Energy-level correlations provide general tools for the statistical 
description of disordered systems, helping in our understanding
of the localization transition. An important statistical measure of
spectral correlations is the two-level correlation function (TLCF)
of the density of states (DOS), which measures the correlations of the
DOS at two different energies. This function has been derived analytically
for the PRBM model in the two limiting cases of weak and strong disorder by
mapping the corresponding Hamiltonian onto an effective $\sigma$ model of a
one-dimensional (1D) nature \cite{ME00} and using renormalization-group methods,
\cite{Le89,Le99} respectively. We stress that, unlike the 1D PRBM model,
it has not until now been possible to analytically solve the most interesting,
from an experimental point of view, disordered models with long-range transfer
terms in $d=2$ and $3$. Thus, explicit results for the TLCF in
two-dimensional (2D) and three-dimensional (3D)
models are still lacking, and finding this function is essential in order to
fully understand the MITs. For this reason, we addressed the problem using
numerical calculations. An important and closely related quantity, the
level-number variance (LNV), will be also considered.

In this work we numerically calculate the TLCF and the LNV of critical
$d$-dimensional random matrix ensembles with long-range off-diagonal
elements and orthogonal or unitary symmetry. Since MITs generically take
place at strong disorder (conventional Anderson transition, quantum Hall
transition, transition in $d=2$ for electrons with strong spin-orbit coupling,
etc.), we will restrict ourselves to the study of both quantities in this regime.
In the 1D case our results for the TLCF are in good agreement, except for the
numerical coefficient, with existing analytical estimates. \cite{ME00} In addition,
we propose expressions for the TLCF in the $d=2$ and $3$ cases. Apart from the
importance of these findings from a general point of view, they may be relevant
for several real physical systems (see Sec. \ref{mam}).

The paper begins by first describing the model and the methods used for the
calculations in Sec. \ref{mam}. The results for the TLCF and the LNV in models
with orthogonal or unitary symmetry are presented in Secs. \ref{res} and \ref{lnv},
respectively. Finally, Sec. \ref{sum} summarizes our findings.

\section{Model and methods}
\label{mam}

In order to fully represent the mesoscopic systems we introduce an
explicit dependence on dimensionality $d$ in the widely studied
PRBM ensemble. \cite{Mi00,CO02,EM00,KT00,MF96,ME00,VB00,Va02,CG01,%
Cu02,YK03,YK03a,Cu03b,Ga03,Cu03,Cu03a} Thus, we consider a generalization
to $d$ dimensions of this ensemble. The corresponding Hamiltonian, which
describes noninteracting electrons on a disordered $d$-dimensional square
lattice with random long-range hopping, is represented by random Hermitian
$L^d \times L^d$ matrices $\widehat{{\cal H}}$ (real for $\beta=1$
or complex for $\beta=2$), whose entries are randomly drawn from a normal
distribution with zero mean, $\left\langle {\cal H}_{ij} \right\rangle =0$,
and a variance that depends on the distance between the lattice sites
$\boldsymbol{r}_i$
\begin{equation}
\left\langle |{\cal H}_{ij}|^2\right\rangle =\frac{1}
                     {1+(|\boldsymbol{r}_i-\boldsymbol{r}_j|/b)^{2\alpha}}
\times\left\{\begin{array}{ll}
                    \frac {1}{2\beta}     \ ,\quad & i\neq j\\
                    \frac {1}{\beta} \,\ ,\quad & i=j
       \end{array}\right.
\label{h1dor}
\end{equation}
in which standard Gaussian ensemble normalization is used. \cite{BF81}

Using field-theoretical methods, \cite{Mi00,EM00,KM97,KT00,Kr99,Le89,AL97,MF96}
the PRBM model was shown to undergo a sharp transition at $\alpha=d$ from
localized states for $\alpha>d$ to delocalized states for $\alpha<d$. This
transition shows all the key features of the Anderson MIT, such as multifractality
of the eigenfunctions and nontrivial spectral compressibility at criticality.
In what follows, we focus on the critical value $\alpha=d$.

The parameter $b^d$ in Eq. (\ref{h1dor}) is an effective bandwidth that
serves as a continuous control parameter over a whole line of criticality,
i.e, for an exponent equal to $d$ in the hopping elements
${\cal H}_{ij} \sim b^d$. \cite{Le89} Furthermore, it determines the critical
dimensionless conductance in the same way as the dimensionality labels the
different Anderson transitions. Each regime is characterized by its respective
coupling strength, which depends on the ratio
$(\langle |{\cal H}_{ii}|^2 \rangle/\langle |{\cal H}_{ij}|^2 \rangle)^{1/2}
\propto b^{-d}$ between diagonal disorder and the off-diagonal transition
matrix elements of the Hamiltonian. \cite{Ef83}

Many real systems of interest can be described by Hamiltonians (\ref{h1dor}).
Among such systems are optical phonons in disordered dielectric materials
coupled by electric dipole forces, \cite{Yu89} excitations in two-level systems
in glasses interacting via elastic strain, \cite{BL82} magnetic impurities in
metals coupled by an $r^{-3}$ Ruderman-Kittel-Kasuya-Yodida interaction,
\cite{CB93} and impurity quasiparticle states in two-dimensional disordered
$d$-wave superconductors. \cite{BS96} It also describes a particle moving fast
through a lattice of Coulomb scatterers with power-law singularity, \cite{AL97}
the dynamics of two interacting particles in a 1D random potential, \cite{PS97}
and a quantum chaotic billiard with a nonanalytic boundary. \cite{CP99}

The TLCF is defined in the usual way
\begin{equation}
\label{r2s1}
R_{d,\beta}(\omega)={\frac{1}{\langle\nu(\epsilon)\rangle^2}}\langle\nu
(\epsilon+\omega/2)\nu(\epsilon-\omega/2)\rangle\;,
\end{equation}
where $\nu(\epsilon)=L^{-d} \text{Tr}\, \delta(\epsilon-\widehat{{\cal H}})$
is the fluctuating DOS and $\langle \; \rangle$ denotes averaging over disorder
realizations. At the critical point $R_{d,\beta}(\omega)$ acquires a scale-invariant
form, if considered as a function of $s=\omega/\Delta$, the frequency normalized to
the mean level spacing $\Delta=1/L^d\langle\nu(\epsilon)\rangle$. \cite{AZ88,SS93,KL94}
In the case of constant average DOS, $R_{d,\beta}(s=\omega/\Delta)$ can be simply
rewritten as
\begin{equation}
\label{r2s2}
R_{d,\beta}(s)=\delta(s)+\sum_n p(n;s)\;,
\end{equation}
where $p(n;s)$ is the distribution of distances $s_n$ between $n$ other
energy levels and the $\delta(s)$ function describes the self-correlation
of the levels. \cite{Me91}

The strong disorder limit ($b \ll 1$) of the 1D PRBM model can be studied
using the renormalization-group method of Refs.~\onlinecite{Le89} and 
\onlinecite{Le99}. For orthogonal symmetry, the following result is obtained
for the TLCF at the center of the spectral band: \cite{ME00}
\begin{equation}
\label{r2s-1}
R_{1,1}(s)=1+\delta(s)-\text{erfc}\left(a_{1,1}\, \frac{|s|}{b}\right)\; ,
\end{equation}
where $\text{erfc}(x)=(2/\sqrt{\pi})\int_x^\infty\exp(-t^2)dt$ is the
complementary error function and $a_{1,1}=1/\sqrt{\pi}$, whereas
for unitary symmetry
\begin{equation}
\label{r2s-2}
R_{1,2}(s)=1+\delta(s)-\exp \left(-a_{1,2}\, \frac{s^2}{b^2}\right)
\end{equation}
with $a_{1,2}=2/\pi$.
Note that for small $s$, $R_{1,\beta}(s)$ behave as $s^{\beta}$ thus
reflecting the levels repulsion effect, whereas they tend asymptotically to
$1$ at large values of $s$.

Another important quantity, which characterizes fluctuations in the level
density on larger scales than the mean spacing, is the variance
\begin{equation}
\label{s2n}
\Sigma^2_{d,\beta}(\langle n \rangle)=
\langle n^2 \rangle-\langle n \rangle^2
\end{equation}
of the number of levels in an energy window that contains
$1\ll \langle n \rangle \ll N$ on average. This variance is given in terms
of the TLCF by $\Sigma^2_{d,\beta}(\langle n \rangle)=
\int_{-\langle n \rangle}^{\langle n \rangle} (\langle n \rangle
-|s|)R_{d,\beta}(s)\,ds$.
The number variance (\ref{s2n}) is a statistical quantity that provides
a quantitative measure of the long-range rigidity of the energy spectrum.
For the Poisson distribution, the levels are uncorrelated and there are
large level-number fluctuations, leading to a linear variance
$\Sigma^2_{d,\beta}(\langle n \rangle)=\langle n \rangle$. On the other hand,
the level correlations in the Wigner-Dyson statistics make the spectrum more
rigid and the number variance grows only logarithmically
$\Sigma^2_{d,\beta}(\langle n \rangle) \sim \ln \langle n \rangle$.
However, in the critical regime, the variance has been conjectured to be
Poisson-like \cite{AZ88,AK94,AM95}
\begin{equation}
\label{chi0}
\Sigma^2_{d,\beta}(\langle n \rangle) \sim \chi \langle n \rangle\;,
\end{equation}
where the level compressibility $\chi$ is another important parameter for
characterizing the MIT and which takes values $0 \le \chi \le 1$, with zero
referring to delocalized states and unity the localized states.

For the computation of $R_{d,\beta}(s)$ and $\Sigma^2_{d,\beta}(\langle n \rangle)$,
we unfold the spectrum in each case to a constant density and rescale it so as to
have the mean spacing equal to unity. Then we calculate $p(n;s)$ and use
Eqs. (\ref{r2s2}) and (\ref{s2n}), respectively. The system sizes
range between $L=500$ and $6000$ in 1D, $L=20$ and $100$ in 2D, and between
$8$ and $14$ in 3D, whereas  $b^d$ ranges in the interval $0.02 \le b^d \le 0.12$.
We consider a small energy window, containing about 10\% of the states around
the center of the band. The number of random realizations is such that the number
of critical levels included for each $L$ is roughly $1.2\times 10^6$. In order to
reduce edge effects, periodic boundary conditions are included.

\section{Two-level correlation function}
\label{res}

In this section, we numerically compute the TLCF $R_{d,\beta}(s)$ of Hamiltonians
(\ref{h1dor}) with the orthogonal or unitary symmetry for different values of the
inverse coupling constant $b^d \ll 1$ and various system sizes. We also compare
our results with the analytical estimates of Ref.~\onlinecite{ME00} for the 1D
model.

\subsection{Orthogonal symmetry}
\label{ors}

\begin{figure}
\begin{center}
\includegraphics[width=6.0cm]{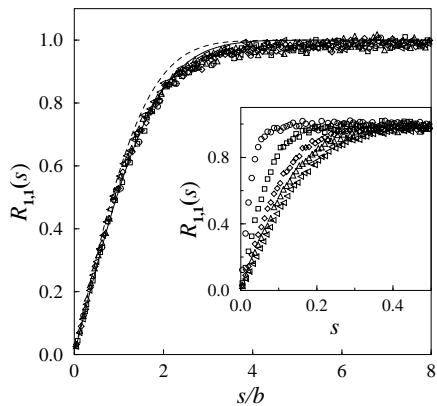}
\caption{\label{fig1}
$R_{1,1}(s)$ for the 1D model (\ref{h1dor}) with orthogonal symmetry
($\beta=1$) as a function of the rescaled variable $s/b$ at different
$b$ values for several system sizes:
$b=0.02, L=4000$ (circles),
$b=0.05, L=6000$ (squares),
$b=0.08, L=1000$ (diamonds),
$b=0.1, L=4000$ (up triangles),
and $b=0.12, L=1000$ (left triangles).
The solid and dashed lines represent the renormalization-group estimate
[Eq. (\ref{r2s-1})], with the fitting parameter $a_{1,1}=0.502$ and the
predicted $a_{1,1}=1/\sqrt{\pi}$, respectively. The inset shows the same
data on the scale $s$.}
\end{center}
\end{figure}

Let us first check the renormalization-group result, Eq. (\ref{r2s-1}),
corresponding to the 1D Hamiltonian (\ref{h1dor}) with orthogonal symmetry
$\beta=1$. The inset of Fig. \ref{fig1} displays our results for $R_{1,1}(s)$
at different $b$ values for several system sizes:
$b=0.02, L=4000$ (circles),
$b=0.05, L=6000$ (squares),
$b=0.08, L=1000$ (diamonds),
$b=0.1, L=4000$ (up triangles),
and $b=0.12, L=1000$ (left triangles).
If the horizontal axis is rescaled by a factor $1/b$, then all data should
collapse onto a single curve.
The main panel of Fig. \ref{fig1} shows $R_{1,1}(s)$ as a function of the
rescaled variable $s/b$, thus, confirming the $s/b$ dependence of $R_{1,1}(s)$.
The best fit of this data set to Eq. (\ref{r2s-1}) gives the fitting
parameter $a_{1,1}=0.502 \pm 0.003$, which is smaller than the predicted value
$1/\sqrt{\pi}=0.564$. Reference~\onlinecite{ME00} reported numerical values of
$R_{1,1}(s)$ at $b=0.1$ for two small system sizes $L=256$ and $512$,
and, as in our calculations, their results were also in relative disagreement
with the value $1/\sqrt{\pi}$ in Eq. (\ref{r2s-1}). The solid and dashed
lines represent Eq. (\ref{r2s-1}), with the fitting parameter $a_{1,1}=0.502$
and the predicted $a_{1,1}=1/\sqrt{\pi}$, respectively. Note that in
Ref.~\onlinecite{ME00} a different normalization was used in Eq. (\ref{h1dor}).

\begin{figure}
\begin{center}
\includegraphics[width=6.0cm]{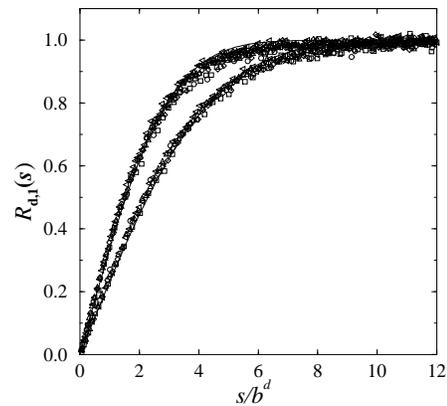}
\caption{\label{fig2}
$R_{d,1}(s)$ for the 2D (upper curve) and 3D (lower curve)
models with orthogonal symmetry $\beta=1$ as a function of the rescaled
variable $s/b^d$ at different $b^d$ values for several system sizes:
$b^2=0.02, L^2=40^2$ (circles),
$b^2=0.05, L^2=20^2$ (squares),
$b^2=0.08, L^2=60^2$ (diamonds),
$b^2=0.1, L^2=100^2$ (up triangles),
$b^2=0.12, L^2=30^2$ (left triangles),
$b^3=0.02, L^3=12^3$ (circles),
$b^3=0.05, L^3=14^3$ (squares),
$b^3=0.08, L^3=12^3$ (diamonds),
$b^3=0.1, L^3=8^3$ (up triangles),
and $b^3=0.12, L^3=10^3$ (left triangles).
The solid lines are fits to Eq. (\ref{r2s-3}), with the fitting
parameters $a_{2,1}=0.308$ and $a_{3,1}=0.208$.}
\end{center}
\end{figure}

Next we consider the $d=2$ and $3$ cases for which, as mentioned in the
Introduction, there are no analytical predictions. The results for $R_{d,1}(s)$
are shown in Fig. \ref{fig2} in which we were able to collapse both sets of data
onto single curves by rescaling the normalized spacing $s$ to the coupling
constant $1/b^d$ of the model. Given the similarity of these results with those
for the 1D model, a curve of the form 
\begin{equation}
\label{r2s-3}
R_{d,1}(s)=1+\delta(s)-\text{erfc}\left(a_{d,1}\, \frac{|s|}{b^d}\right)
\end{equation}
was fitted to the data points in this graph and found that $a_{2,1}=0.308 \pm 0.001$
and $a_{3,1}=0.208 \pm 0.001$. These fits are represented as solid lines in
Fig. \ref{fig2}. Thus, the system dimensionality $d$ of the TLCF enters via the
inverse-coupling constant $b^{-d}$. We stress that Eq. (\ref{r2s-3}) gives a fairly
good fit to the data. The reported data correspond to
$b^2=0.02, L^2=40^2$ (circles),
$b^2=0.05, L^2=20^2$ (squares),
$b^2=0.08, L^2=60^2$ (diamonds),
$b^2=0.1, L^2=100^2$ (up triangles),
$b^2=0.12, L^2=30^2$ (left triangles),
$b^3=0.02, L^3=12^3$ (circles),
$b^3=0.05, L^3=14^3$ (squares),
$b^3=0.08, L^3=12^3$ (diamonds),
$b^3=0.1, L^3=8^3$ (up triangles),
and $b^3=0.12, L^3=10^3$ (left triangles).

These results allow us to generalize the 1D analytical result [Eq. (\ref{r2s-1})]
to the 2D and 3D models by simply replacing the inverse-coupling constant $b$
of the $1d$ case by the corresponding to the $d$-dimensional case $b^d$.

The observed $b^d$ dependence of $R_{d,1}(s)$ in Eq. (\ref{r2s-3})
is not surprising since other critical properties, such as the correlation
dimension $d_2$ in 1D and 2D present a similar behavior toward $b^d$.
Specifically, $d_2=1-1/\pi b$ ($b \gg 1$) and $d_2=2b$  ($b \ll 1$) were
derived in Refs. \onlinecite{MF96} and \onlinecite{Mi00}, whereas
$d_2=2-a_2/b^2$ ($b^2 \gg 1$) and $d_2=c_2b^2$  ($b^2 \ll 1$) were numerically
found in Ref. \onlinecite{Cu03c}.

\subsection{Unitary symmetry}
\label{uns}

\begin{figure}
\begin{center}
\includegraphics[width=6.0cm]{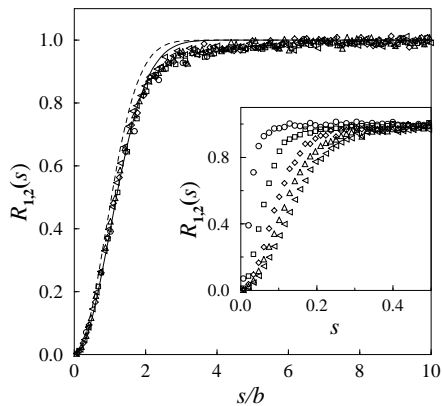}
\caption{\label{fig3} 
As in Fig. 1, for the 1D model (\ref{h1dor}) with unitary symmetry
($\beta=2$). Data correspond to 
$b=0.02, L=500$ (circles),
$b=0.05, L=1000$ (squares),
$b=0.08, L=2000$ (diamonds),
$b=0.1, L=500$ (up triangles),
and $b=0.12, L=1000$ (left triangles).
The solid and dashed lines represent Eq. (\ref{r2s-2}), with
the fitting parameter $a_{1,2}=0.495$ and the predicted $a_{1,2}=2/\pi$,
respectively. The inset shows the same data on the scale $s$.}
\end{center}
\end{figure}

As in Sec. \ref{ors} we first analyze the 1D case in order to compare
the numerical data with the analytical result [Eq. (\ref{r2s-2})].
Figure \ref{fig3} shows $R_{1,2}(s)$ as a function of the rescaled variable
$s/b$. As expected, all data points collapse onto the same curve. The inset
displays the same data on the $s$ scale. The values of $b$ and $L$ reported
are:
$b=0.02, L=500$ (circles),
$b=0.05, L=1000$ (squares),
$b=0.08, L=2000$ (diamonds),
$b=0.1, L=500$ (up triangles),
and $b=0.12, L=1000$ (left triangles).
Fitting these data to Eq. (\ref{r2s-2}) gives $a_{1,2}=0.495 \pm 0.005$,
which again is small that the predicted value $2/\pi=0.637$. To our
knowledge, this is the first numerical confirmation of Eq. (\ref{r2s-2}). 
The solid and dashed lines represents Eq. (\ref{r2s-2}), with the fitting
parameter $a_{1,2}=0.495$ and the predicted $a_{1,2}=2/\pi$, respectively.

\begin{figure}
\begin{center}
\includegraphics[width=6.0cm]{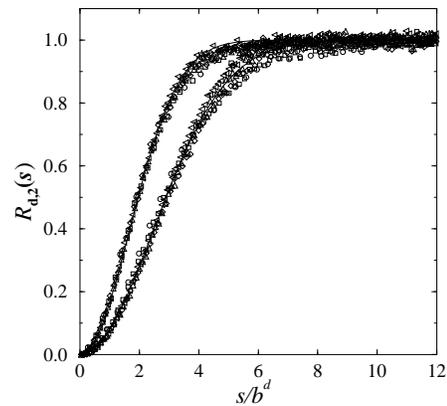}
\caption{\label{fig4}
As in Fig. 2, for the 2D (upper curve) and 3D (lower curve)
models with unitary symmetry ($\beta=2$). Data correspond to
$b^2=0.02, L^2=40^2$ (circles),
$b^2=0.05, L^2=20^2$ (squares),
$b^2=0.08, L^2=60^2$ (diamonds),
$b^2=0.1, L^2=80^2$ (up triangles),
$b^2=0.12, L^2=30^2$ (left triangles),
$b^3=0.03, L^3=12^3$ (circles),
$b^3=0.05, L^3=14^3$ (squares),
$b^3=0.08, L^3=10^3$ (diamonds),
$b^3=0.1, L^3=10^3$ (up triangles),
and $b^3=0.12, L^3=8^3$ (left triangles).
The solid lines are fits to Eq. (\ref{r2s-3}), with the fitting
parameter $a_{2,2}=0.181$ and $a_{3,2}=0.083$, respectively.}
\end{center}
\end{figure}

The results for $R_{d,2}(s)$ in the $d=2$ and $3$ models are shown in
Fig. \ref{fig4}, in which one can clearly appreciate the collapse of both
sets of data when represented as a function of the rescaled variable $s/b^d$. 
As in the cases of $\beta=1$, a curve of the form 
\begin{equation}
\label{r2s-4}
R_{d,2}(s)=1+\delta(s)-\exp \left(-a_{d,2}\, \frac{s^2}{b^{2d}}\right)
\end{equation}
was fitted to the data points in this graph, where $a_{d,\beta}$ is a 
fitting parameter. The data reported correspond to
$b^2=0.02, L^2=40^2$ (circles),
$b^2=0.05, L^2=20^2$ (squares),
$b^2=0.08, L^2=60^2$ (diamonds),
$b^2=0.1, L^2=80^2$ (up triangles),
$b^2=0.12, L^2=30^2$ (left triangles),
$b^3=0.03, L^3=12^3$ (circles),
$b^3=0.05, L^3=14^3$ (squares),
$b^3=0.08, L^3=10^3$ (diamonds),
$b^3=0.1, L^3=10^3$ (up triangles),
and $b^3=0.12, L^3=8^3$ (left triangles).
The solid lines in Fig. \ref{fig4} are fits to Eq. (\ref{r2s-4})
with fitting parameters $a_{2,2}=0.181 \pm 0.001$ and $a_{3,2}=0.083 \pm 0.001$.
Note that this equation gives a fairly good fit to the data.

For the system sizes considered ($L^d \gg 1$) we have checked that
$R_{d,\beta}(s)$ is an $L$-independent universal scale-invariant
function, thus confirming the existence of a critical distribution
exactly at the transition. Furthermore, using the nearest-level
distribution $p(0;s)$, we verified that the normalized nearest
level variances are indeed scale invariant at each critical point
studied.\cite{Cu99}

Before concluding the section, we summarize in Table~\ref{table1} the
nonuniversal constants $a_{d,\beta}$ found. Note that each value is
different for every $d$ and $\beta$, thus reflecting its dependence on
the Hamiltonian symmetry and dimensionality. The values in brackets were
obtained from the spectral compressibility (see Sec. \ref{lnv}).

\begin{table}[b]
\caption{\label{table1}Nonuniversal constants $a_{d,\beta}$ of the TLCF.
The values in brackets were obtained from the spectral compressibility.}
\begin{ruledtabular}
\begin{tabular}{|r|ccc|}
& $d=1$   & 2 & 3\\
\hline
$\beta=1$ & 0.502 (0.497) & 0.308 (0.291) & 0.208 (0.232) \\
$2$       & 0.495 (0.483) & 0.181 (0.206) & 0.083 (0.105)\\
\end{tabular}
\end{ruledtabular}
\end{table}

\section{Level number variance}
\label{lnv}

This section is devoted to the calculation of the LNV
$\Sigma^2_{d,\beta}(\langle n \rangle)$ and the spectral compressibility
$\chi$ of models (\ref{h1dor}) with the orthogonal or unitary symmetry in
the strong coupling regimen, $b^d \ll 1$. Our results for $\chi$
are compared with the analytical prediction of Ref.~\onlinecite{CK96}.

It is well known that the statistical properties of spectra of disordered
one-electron systems are closely related to the localization properties of
the corresponding wave functions. \cite{SS93,KL94,AS86} In particular,
Ref. \onlinecite{CK96}, based on a Brownian motion level approximation
combined with an assumption concerning the decoupling of the energy levels
and eigenfunction correlations, derived the following relation for multifractal
eigenstates 
\begin{equation}
\label{chivd2}
\chi=\frac{1}{2} \left( 1-\frac{d_2}{d} \right)\,,
\end{equation}
where $d_2$ is the correlation dimension of the critical wave functions.
According to this result, the spectral compressibility $\chi$ should tend
to $1/2$ at the limit of very sparse eigenstates $d_2 \to 0$, and not
to the Poisson value $\chi=1$. Although Eq. (\ref{chivd2}) has been widely
confirmed at weak multifractality (weak coupling limit) its validity at
strong coupling (small $d_2$) raises many doubts. More precisely, in
Ref. \onlinecite{ME00}, it was analytically shown that at the limit of small
$b$, $\chi \to 1$ for the one-dimensional model (\ref{h1dor}). This tendency
has been also numerically demonstrated in Refs. \onlinecite{ME00} and
\onlinecite{NK03}, and for the Anderson model in $d \ge 4$.\cite{ZK98}

We are therefore especially interested in the calculation of $\chi$ for $d=2$
and $3$ in order to check whether or not Eq. (\ref{chivd2}) adequately
describes its behavior at strong coupling $b^d \ll 1$. The compressibility
$\chi$ can be expressed through the TLCF as
\begin{equation}
\label{chivs}
\chi=\int_{-\infty}^{\infty} ds\,[R_{d,\beta}(s)-1]\,.
\end{equation}
Substitution of Eqs. (\ref{r2s-1}), (\ref{r2s-2}), (\ref{r2s-3}), and
(\ref{r2s-4}) into Eq. (\ref{chivs}) yields the spectral compressibility
for orthogonal ($\beta=1$) as well as for unitary symmetry ($\beta=2$)
\begin{equation}
\label{chi2d}
\chi=1-c_{d,\beta}\,b^d\,,
\end{equation}
where $c_{d,1}=2/a_{d,1}\sqrt{\pi}$ and $c_{d,2}=\sqrt{\pi/a_{d,2}}$.
Equation (\ref{chi2d}) constitutes a generalization to $d$ dimensions of
the 1D analytical estimates of Ref. \onlinecite{ME00}. Notice that at
the limit of very strong coupling $b^d \to 0$, $\chi$ tends to the Poisson
value $1$ in disagreement with Eq. (\ref{chivd2}).

An alternative way to directly calculate $\chi$ is from the asymptotic
behavior of the LNV [Eq. (\ref{chi0})]. We will show that for $b^d \ll 1$,
relation (\ref{chi2d}) is satisfied, thus, giving further support to the
proposed relations for the TLCF [Eqs. (\ref{r2s-3}) and (\ref{r2s-4})].

\begin{figure}
\begin{center}
\includegraphics[width=6.0cm]{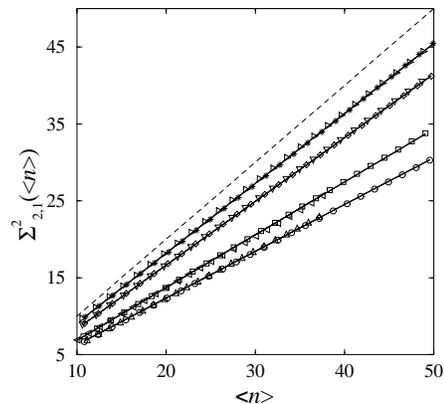}
\caption{\label{fig5}
Level number variance $\Sigma^2_{2,1}(\langle n \rangle)$ for the 2D
orthogonal ensemble ($\beta=1$) at different disorders and system sizes: 
$b^2=0.02, L^2=40^2$ (right triangles),
$b^2=0.02, L^2=60^2$ (stars),
$b^2=0.04, L^2=40^2$ (down triangles),
$b^2=0.04, L^2=60^2$ (diamonds),
$b^2=0.08, L^2=20^2$ (left triangles),
$b^2=0.08, L^2=60^2$ (squares),
$b^2=0.1, L^2=20^2$ (up triangles),
and $b^2=0.1, L^2=100^2$ (circles).
Solid lines are fits to Eq. (\ref{chi0}), and the dashed line corresponds
to the Poisson result.}
\end{center}
\end{figure}

\begin{figure}[b]
\begin{center}
\includegraphics[width=6.0cm]{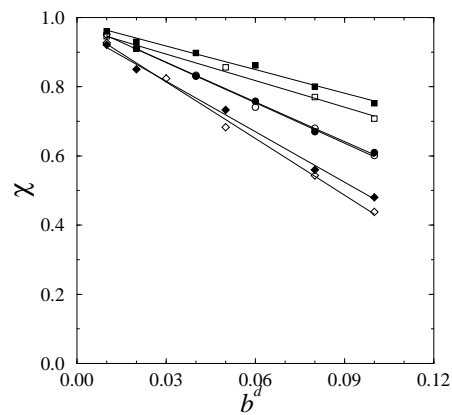}
\caption{\label{fig6}
$b^d$ dependence of the spectral compressibility $\chi$ for the 1D
(squares), 2D (circles), and 3D (diamonds) models (\ref{h1dor})
with orthogonal (solid symbols) or unitary symmetry (open symbols).
Solid lines are fits to the form $\chi=\chi_0-c_{d,\beta}b^d$.}
\end{center}
\end{figure}

In Fig. \ref{fig5}, we show the computed $\Sigma^2_{2,1}(\langle n \rangle)$
for the 2D orthogonal ensemble ($\beta=1$) at different disorders and system
sizes: 
$b^2=0.02, L^2=40^2$ (right triangles),
$b^2=0.02, L^2=60^2$ (stars),
$b^2=0.04, L^2=40^2$ (down triangles),
$b^2=0.04, L^2=60^2$ (diamonds),
$b^2=0.08, L^2=20^2$ (left triangles),
$b^2=0.08, L^2=60^2$ (squares),
$b^2=0.1, L^2=20^2$ (up triangles),
and $b^2=0.1, L^2=100^2$ (circles).
Note that, for each value of $b^2$, $\Sigma^2_{2,1}(\langle n \rangle)$
is $L$ independent, which is a sign of criticality. There is a clear gradual
tendency in the large $\langle n \rangle$ region of
$\Sigma^2_{2,1}(\langle n \rangle)$ toward the Poisson limiting result (dashed
line) as the inverse coupling constant $b^2$ of the model is decreased. We have
checked that for
$\beta=2$ and for the 3D case the behavior is quite similar. The straight lines,
whose slopes correspond to the values of $\chi$, are fits to Eq. (\ref{chi0}) with
the fitting parameters summarized in Fig. \ref{fig6}.

The $b^d$ dependence of the spectral compressibility $\chi$, as obtained from
the previous fits, for the 1D (squares), 2D (circles), and 3D (diamonds)
model (\ref{h1dor}) with orthogonal (solid symbols) or unitary symmetry (open
symbols) is depicted in Fig. \ref{fig6}. This clearly shows that for almost all
values of $b^d$ reported, $\chi$ is greater than the maximum value of $0.5$
predicted by Eq. (\ref{chivd2}). The solid lines are fits to
the form $\chi=\chi_0-c_{d,\beta}b^d$ with the fitting parameters
$c_{1,1}=2.27 \pm 0.14$, $c_{1,2}=2.55 \pm 0.13$, $c_{2,1}=3.88 \pm 0.12$,
$c_{2,2}=3.90 \pm 0.26$, $c_{3,1}=4.87 \pm 0.19$, and $c_{3,2}=5.46 \pm 0.20$.
Using these values of $c_{d,\beta}$ and taking into account their relation
with the nonuniversal constants $a_{d,\beta}$ of the TLCF,
$c_{d,1}=2/a_{d,1}\sqrt{\pi}$ and $c_{d,2}=\sqrt{\pi/a_{d,2}}$, we can obtain
$a_{d,\beta}$ from a different quantity. The corresponding values are given in
brackets in Table \ref{table1}, in which one can appreciate the good agreement
with those obtained from the TLCF. Based on these results, one can say that, in
the case of strong coupling, the Brownian motion level approximation breaks down
and Eq. (\ref{chivd2}) becomes invalid for all values of $d$ considered.

\section{Summary}
\label{sum}

We present the first numerical results for the two-level correlation function
$R_{d,\beta}(s)$ of noninteracting electrons on a $d$-dimensional disordered
system with long-range transfer terms. Models with orthogonal or unitary symmetry
at small values of the inverse-coupling constant $b^{d}$ have been considered.
The 1D analytical results [Eqs. (\ref{r2s-1}) and (\ref{r2s-2})] are confirmed
(except for the numerical constants). We also found that the 1D formulas are
valid for the 2D and 3D models if the inverse-coupling constant $b$ is replaced
by the corresponding to the $d$-dimensional case $b^d$. Another important result
provided by our numerical calculations is that the spectral compressibility, which
is found to be close to $1$ in the limit $b^d \to 0$, does not satisfy the
relation (\ref{chivd2}). The proposed Eqs. (\ref{r2s-3}) and (\ref{r2s-4})
are based on numerical results and, at present, should be considered as conjectural.
So, further analytical work is needed to check these forms of the TLCF and their
origin in the model (\ref{h1dor}).

\begin{acknowledgments}
The author thanks the FEDER and the Spanish DGI for financial support through
Project Nos. BFM2003-03800 and FIS2004-03117.
\end{acknowledgments}

\end{document}